**Raman Detection Threshold Measurements for Acetic Acid in Martian Regolith Simulant JSC-1 in the Presence of Hydrated Metallic Sulfates**

Keith Andrew[1, 3,6], Kristopher A. Andrew[4, 5], Melinda Thomas[1,4,6], Alicia Pesterfield[2], Quentin Lineberry[6], Eric V. Steinfelds[1,6]

[1]Department of Physics and Astronomy, [2]Department of Chemistry, [3]Institute for Astrophysics and Space Science, [4]Applied Physics Institute, Western Kentucky University, Bowling Green, KY 42101, [5]Department of Physics and Astronomy, University of Kentucky, Lexington, KY 40506, [6]Vaon LLC, Pic Laboratory, Franklin, KY 42134

## Abstract

Several measurements, including the data from the laser spectrometer in the Sample Analysis at Mars instrument suite on the Curiosity Rover in Gale crater, have measured seasonal variations in atmospheric methane at the 0.7 ppbv (parts per billion volume) level. As a result, models have been proposed to understand the methane production including novel chemical, geological, meteorological, and biological mechanisms. Biological models often rely on methanogenic extremophile archaea which might be similar to some Earth based organisms which can be studied exhaustively in a laboratory setting. Such organisms might thrive in a subsurface ecosystem involving water and methane and as such could leave a unique biosignature in the Martian regolith that could be preserved over extended periods of time. The resulting mixture of carboxyl, acetyl and hydroxyl groups blended with the metallic sulfates and regolith constituents could produce a Raman signature detectable on the next rover mission if the concentration is high enough. Here we measure the 532 nm and the 780 nm Raman spectra of a variety of molar concentrations of several hydrated metallic sulfates mixed with JSC-1 Martian regolith simulant to identify the signature concentration thresholds for the isolated C=O Raman peak from acetic acid. We find a Raman peak removed from most fragmented molecular Raman bands near 1608 $cm^{-1}$ using the 532 nm laser for the Fe and Mg sulfates to yield a threshold in the 120 - 160 ppmv range for a cutoff S/N of 2.

## I. Introduction

The early detection of periodic releases of methane as atmospheric plumes at select latitudes in the northern hemisphere of Mars in 2003, as seen from both Earth based observations[1,2] and from orbiting sensors,[3,4] indicating that single plume masses on the order of 19 metric ktons of methane had been generated has produced a considerable amount of excitement and research. Estimates of concentrations as high as 30 ppbv (parts per billion volume), even when distributed uniformly throughout the atmosphere, required unusual decay and source channels to match the low abundances found elsewhere. Sources for methanogenesis can be geochemical- most notably low temperature geothermal serpentinization[5], can be from meteoric or cometary atmospheric in-fall, or can be biogenic in origin. However, detailed modeling[6] and an examination of the telluric lines[7] that interfere with the possible methane production lines indicate that these observations of large amounts of methane in the Martian atmosphere are overestimates. In 2015 data from the tunable laser spectrometer of the Sample Analysis at Mars (SAM) instrument suite on Curiosity Rover at Gale crater, after a twenty-month collection window, indicated the presence of varying methane quantities lower than the original estimates to be about 0.69 ± 0.25 ppbv with some periods of higher production near 7.2 ± 2.1 ppbv[8]. By 2017 the SAM suite team had conducted a complete methane isotopologue study including foreign broadening- from carbon dioxide and He, along with molecular rotational and vibrational analysis[9] from the Curiosity tunable laser system that continued to support the observation of these methane levels.  Methane has also been measured in six of the Martian meteorites that have been found on Earth by using a Crush Fast Scan Mass Spectrometer whereby gasses in the meteorites were released and measured and compared to Earth based and carbonaceous chondrite concentrations[10].  Rover measurements show that the Martian surface contains more olivine and pyroxene than found on Earth and in a low temperature aqueous environment can experience serpentization[11] to produce hydrogen and methane in combined processes such as:

$$18Mg_2SiO_4 + 6Fe_2SiO_4 + 26H_2O + CO_2 \rightarrow 12Mg_3Si_2O_5(OH)_4 + 4Fe_3O_4 + CH_4 \qquad (1)$$

providing an abiotic origin for methane and a mechanism for the creation of a subsurface methane rich habitat.  The presence of hydrated olivine rocks provides a natural pathway to methane production via serpentization, that could be catalyzed by volcanic outgassing, that could have been very active during the Noachian period and resulted in underground pockets of stored methane.  Both the low temperatures and pressures at the surface would have been favorable for methane production that could then be released through surface via geothermal activity.

In addition to the direct atmospheric studies data from the Spirit rover in Gusev crater[12], the Opportunity rover in Meridian Planum[13] and Curiosity rover in Gale crater determined the mineralogical composition of the Martian regolith[14].  The dominant minerals are olivine, pyroxene, plagioclase and magnetite with admixtures of monohydrated and polyhydrated sulfates, most notably with Ca, Ni, Mg[15] and Fe[16,17].  In addition, chlorates and perchlorates[18] have been found indicating a harsh environment[19] for any organic compounds or potential extremophile archaea[20]. The surface minerals experience a significant amount of UV flux on Mars, mainly at the shorter wavelengths of UVC (100-280 nm) and of UVB (280- 315 nm). Various works on the biological effects of UV radiation[21] have established that even the present-day instantaneous Martian UV flux would not in itself prevent the existence of life on Mars. However, it has been found that UV initiated photochemical reactions involving perchlorates[22], peroxides and iron oxides enhance the cell death rate of *Bacillus subtilis* by a factor of 10.8 for a sixty second exposure leading to nearly complete bacteriocide[23]. Thus, the current surface environment is photo-chemically uninhabitable to many known species and would work to eradicate contaminating species from landing vehicles from Earth[24] thereby significantly lowering contamination probability as described by the Coleman-Sagan equation[25]. As such, the preservation of a biosignature on the surface of Mars might focus on the subsurface ecosystem[26]. More recent data from perchlorate, oxide, and sulfate measurements has been used to produce Martian simulants for active study on various mechanical[27] and geochemical properties possessed by the surface materials. In addition, significant amounts of frozen water have been found in exposed scarps 1-2 meters below the surface[28] at the mid-latitude regions. The ice cliffs extend for more than 100 m below the surface and may be indicative of a long period of successive snow deposits during a high obliquity period.  Both geochemical and biochemical processes could take advantage of these long-lived ice sheets.

To prepare for future spectral measurements on the surface of Mars researchers are expanding their analysis to include the advantages of Raman spectroscopy with lab based tests on Martian simulants to identify threshold identification values for various molecules. Studies have been done on the sensitivity of the Raman signal for samples in Martian simulant with hydrated salts[29,30,31] chlorates and perchlorates[32], and organic molecules[33] such as chlorophyll[34], carotene[35] in cyanobacteria[36], and a host of reasonable methanogenic based organic molecules[37]. As indicated by high resolution chromatographic analysis of the Murchison meteorite[38] and other carbonaceous chondrites[39] which contain numerous organic molecules[40], including acetic acid and amino acids such as glycine, D-Alanine and L-Alanine, it is clear that many interesting molecules can have a non-biotic Martian origin. Careful modeling of Fickian and Knudsian diffusion at depths beyond 1.0 km indicate that diffusive mechanisms alone cannot account for all of the methane production[41]. Recently an, in principle, testable mass conservation column flux model to describe the sub-annual Gale crater variation of methane production[42] has provided insight on developing some novel approaches to the observed time variations in atmospheric methane concertation[43] on Mars. The authors examine three near surface interaction scenarios for methane production: in the first case the surface adsorbs methane when dry and then releases it upon deliquescence with surface water diffusion and a concentration near $4x10^{-6}$ kg/m$^3$, in the second microorganisms produce the methane, and in the third they examine possible deep subsurface aquifers to produce methane outgassing.  They consider the conservation equation with relevant sources where we have added the UV- regolith interaction surface term to account for buildup from Interplanetary Dust Particles[44], IDP, with photolytic release mechanisms, a serpentization thermal outgassing term effective to depths of up to z meters where subsurface studies and analysis has been done up to 6 km[45], we use these two terms as potential background methane production terms without adding an explicit relaxation term and we include a volcanic outgassing term for release through porous solidified crustal magma giving a production equation of the form:

$$\frac{dM_{CH_4}}{dt} = F_{serpentization} + F_{Deliquescence} + F_{Methanogenic} + F_{UV} + F_{VolGs} \tag{2}$$

$$\frac{dM_{CH_4}}{dt} = \frac{d}{dt}\int \rho_S \gamma_{CH_4} M_{CH_4} A_S \xi_{ser} dz + \frac{d}{dt}\int \rho_s \gamma_{CH_4} M_{CH_4} A_s \theta_{CH_4} dz + \frac{C_{ali}\rho_s}{t_o}\int Q^{\alpha} dz + \int \frac{2\chi D f M_{CH_4}}{3m_w f_{UV}(aT-b)} dA + \int \beta \sum_i R_{Magma_i} \eta_i X_i^l(z) dz$$

$$\xi_{ser} = \frac{K_{eq} n_{CH_4}}{1 + \sum_{j=1}^{N} K_j n_j} \qquad \theta_X = \frac{K_{eq} n_{CH_4}}{1 + K_{eq-CH_4} n_{CH_4} + K_{eq-CO_2} n_{CO_2}} \qquad K_{eq} = \frac{v_{th} h \exp(E_a / RT)}{4\gamma_{CH_4} k_b T} \qquad \alpha = \frac{T - 273.15}{10} \qquad R_{Magma} = \frac{dM_c}{dt}$$

where the total serpentization process up to 10 km below the surface can be represented by a weakly coupled Langmuir interacting product term $\xi_{Tot}$ with indexed equilibrium constants for each interaction and associated number densities $n_j$,  for each activation energy $E_a$, the chemical kinetics can be altered to include strongly interacting BET or Freundlich interactions to accommodate potential multilayering and surface roughness factors, the first deliquescence term corresponds to a Langmuir isotherm adsorption reaction in the soil, the second term corresponds to the production by methanogenic species where, $\rho_s$ is the soil density,  $\gamma$ is the monolayer coverage of methane per unit surface area, $A_s$ is the specific surface area, $\theta$ is the fractional coverage ratio by methane or species j, $K_{eq}$ is the chemical reaction equilibrium constant for methane or species j, $n_x$ is the number density of chemical species x, $v_{th}$ is the thermal velocity of the chemical species, $E_a$ is the adsorption energy, $C_{ali}$ is the soil content of aliphatic hydrocarbons expressed as a dimensionless ratio in kg/kg of soil, $t_o$ is the baseline soil residence time for carbon at 273 K, Q is a constant chosen to be indicative of typical biological methane emission where they find the value near 2 is consistent with terrestrial studies, h is the Planck constant, R is the gas constant, $k_b$ is the Boltzmann constant, where each term is integrated to a depth z. The surface adsorption interaction is treated as a Langmuir interaction with a deliquescent interaction with water and would generally require more energy than has been seen in the lab environment, near 36 kJ $mole^{-1}$, the methanogenic model requires microorganisms to produce the methane, and their final model examines the existence of deep subsurface aquifers as a source of methane. The $F_{uv}$ term accounts for the solar ultraviolet surface release of methane[46] related to sedimenting interplanetary dust particles, IDP, measured by the Rover Environmental Monitoring Station on the Mars Science laboratory[47]. Production rates for these interactions are measured to be about $8x10^{-4}$ ppbv $sol^{-1}$ and depend upon the radiant UV intensity, averaged over 200-400 nm, the mass fraction of carbon in the regolith $\chi$, surface roughness characterized by the diameter of the regolith grain D, the fraction of in-falling carbon that is accreted f, the molecular weight of the carbon containing molecule $m_w$, and the overall quantum efficiency of the photolytic conversion process as a function of temperature expressed as (a T - b) where a=2.76 x $10^{-15}$ mole/JK and b = 1.54 x $10^{-13}$ mole/J, integrated over the total the total surface area exposed. For well mixed fine-grained regolith this last term produces up to 11 ppbv with variations that depend on latitude and seasonal temperatures, grain size and weather distributions and expected fractional conversion rates. The $F_{VolGas}$ term is for methane produced by volcanic outgassing[48] on Mars. The Martian magma production rate[49] is given by $R_{Magma}$, for $M_{crit}$ which is the mass of extracted liquid magma, $X^l_{melt}$ (z) is the melt concentration[50] of the extracted species in the melt which depends upon the depth below the surface, and $\eta$ is the overall efficiency of the process where values are estimated from the Antarctic Mars meteorites and the Martian surface magnetic fields with magnitudes near 4 nT. For the currently solidified upper crust where up to 25% of the surface is covered by volcanic magma predominately[51] in the Tharsis and Syrtis regions and a volume on the order of 12 x $10^6$ $km^3$ with an internal geothermally active model[52] of the core outgassing can continue at rates that are dependent upon the porosity, depth of the crust and includes $CO_2$ and $H_2O$ components which contribute to significant atmospheric gas production, can stably be productive since the Noachian period and can be deeper than 1 km below the surface, where these gasses can serve as precursors to later time serpentization. The parametrization of Eq. (2) is such that there are many solutions that yield reasonable methane values without need for extraordinary events or circumstances, better measurements will produce more robust constraints allowing for the determination of the dominant terms responsible for producing the observed methane on Mars. A number of time scales arise in this model: the volcanic outgassing through porosity and the associated production of $CO_2$ and $H_2O$ that can participate in serpentization can occur over time scales on the order of 100 million Earth years, the serpentization and aquifer formation can be tens of millions of years, the methanogenic production of methane can have a daily, annual, decadal or longer timeframe, the surface deliquescent interactions and production can occur on a daily and seasonal time scale and the in-fall of IPD occurs over the entire age of the planet and the pyrolysis surface UV reaction occurs in less than a second.

Initial data and boundary conditions for implementing a column model can be extracted from the rover/satellite database and from General Circulation Models such as the open source Mars Climate Database v5.3, MCD[53,54]. In this case the column model serves as a high-resolution test of local conditions with inputs from the global circulation model. The MCD is a database of meteorological fields derived from General Circulation Model (GCM) numerical simulations of the Martian atmosphere and validated using available observational data. The MCD includes complementary post-processing schemes such as high spatial resolution interpolation of environmental data and means of reconstructing the variability thereof. Global maps can be constructed that include variables such as temperature, pressure, water content, dust content[55], wind speed, fractional gas content[56] such as CO or $CO_2$, surface ice, etc. In Fig. (1) we show a sample run for surface pressure in Pascals, and temperature in Kelvins, water vapor for a column in kg/m$^2$, solar flux at surface in W/m$^2$, water vapor volume mixing ratio in mol/mol and water vapor column in kg/m$^2$.

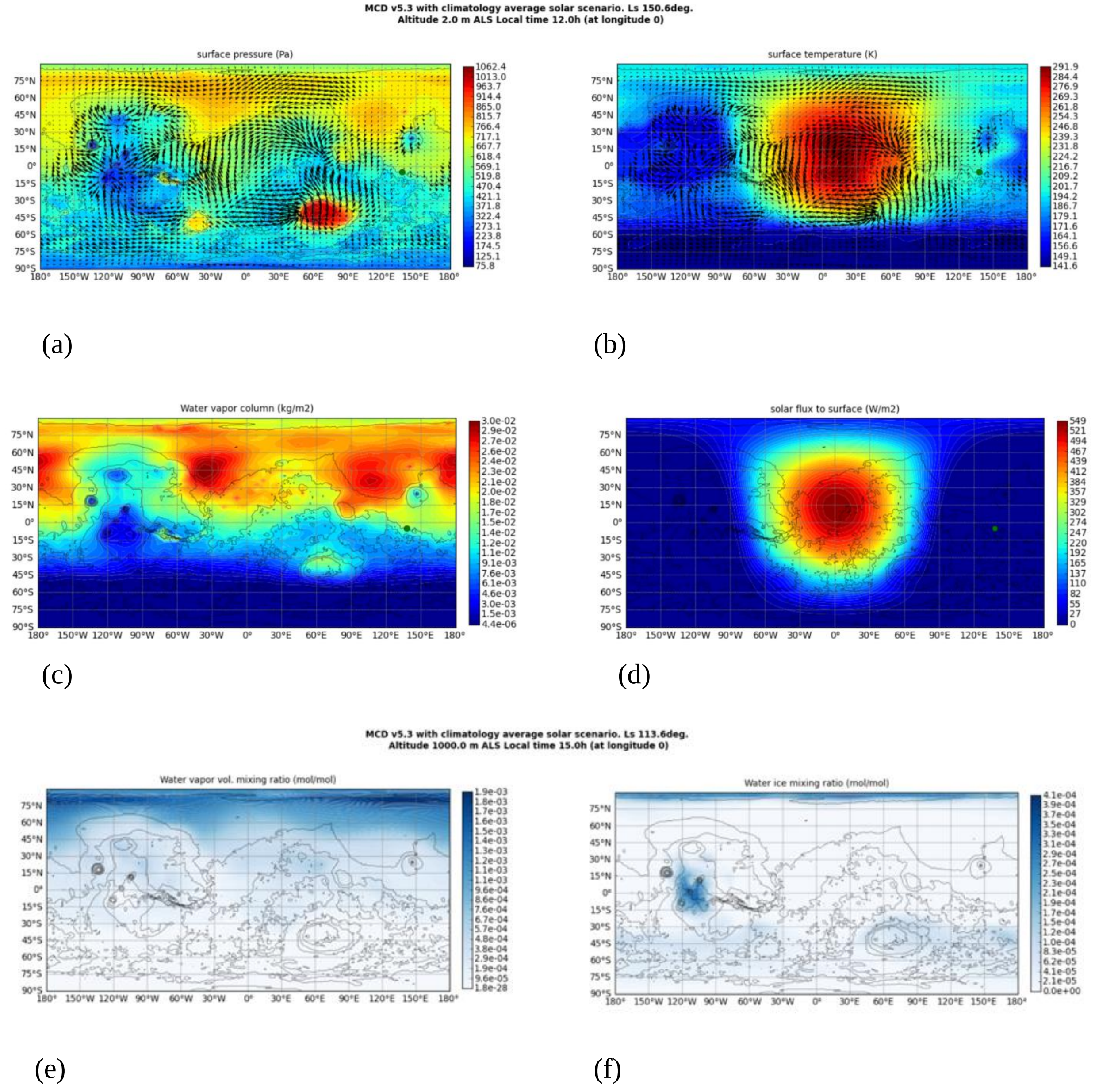

Fig. 1 Mars Climate Database v5.3 output for a Global Circulation Model showing (a) surface pressure in Pascals with vectorize wind speeds at a height of 2.0 m, (b) surface temperature in Kelvins with vectorized wind speeds at a height of 2.0 m, (c) variations in water vapor column values in kg/m$^2$, (d) variations in solar flux to the surface in W/m$^2$ with the sun centered at noon in middle of the map, (e) the global water[57] vapor mixing ration in mol/mol, and (f) the water ice mixing ratio mol/mol.

Each of the global maps can be displayed for altitudes near the surface, at the troposphere, at the boundary layer[58], at the mesosphere, or at the thermosphere and can highlight phenomena most related to the winds, weather, atmosphere, water cycle, chemistry, glaciology, landing engineering, surface meteorology and radiative balance. Data can also be collected for a fixed location, such as the Curiosity region in Gale crater. For a fixed latitude and longitude, the local variation of these parameters can be determined as functions of time or altitude as shown in Fig. (2).

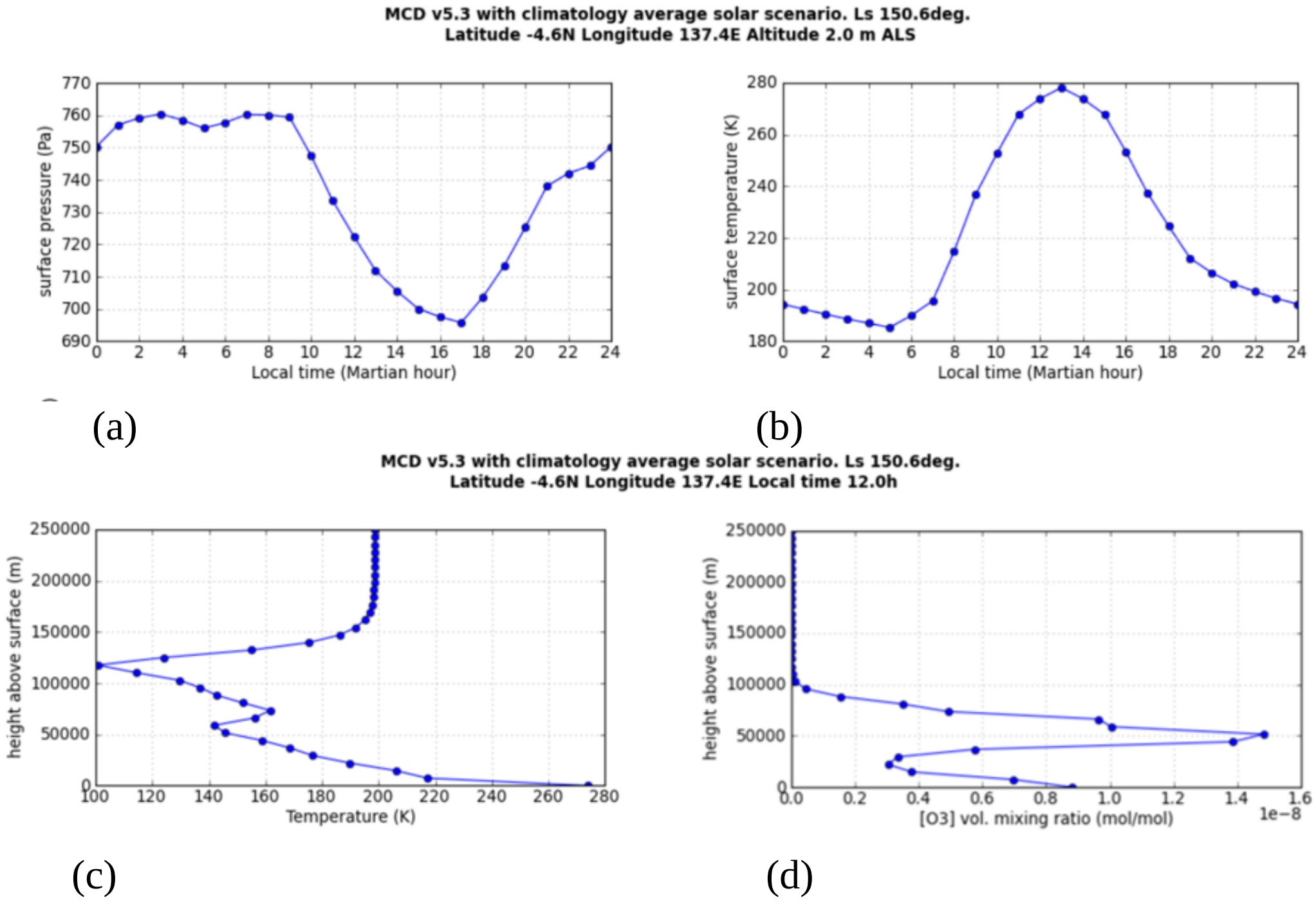

Figure (2) Gale crater location showing changes in: (a) surface pressure on Pascals over one sol, (b) surface temperature in Kelvins for one sol, (c) atmospheric temperature as a function of altitude up to 250 km[59] and (d) variation in Ozone volume mixing ratio as a function of altitude[60].

The range of parameters from the GCM provide a significant base for modeling and as input values for Eq. (2). The associated ice, depth, radiation, dust and estimated magma flow regions and layering provide a reasonable set of values to initiate a general study of the model.

Another approach to investigating potential organic molecules[61] is to identify those that might be similar to Earth based methanogenic extremophiles that take advantage of the observed methane signature at Mars.[62] The metabolic requirements of methanogenic archaea are in principle compatible with these extreme environmental conditions[63] especially the methanogenic permafrost and ice related psychrophiles in the cryobiosphere[64]. Several common extremophiles and their average ecosystem parameters are listed below in Table 1.

Parameters for Methanogenic Extremophiles

| Methanogenic Archea[65] | *M. formicicum*[66] | *M. bryantii*[67] | *M. ivanovii*[68] | *M. alcaliphilum*[69] | *M. espanolae*[70] | *M. palustre*[71] | *M. uliginosum*[72] | *M. subterraneum*[73] |
|---|---|---|---|---|---|---|---|---|
| Morphology and structure | Rod:(0.4-0.8) x(2-15), filaments, clumps | Rod:(0.5-1)x(1.5) filaments, clumps | Rod: 0.5-0.8x1.2 Filaments | Rod: 0.5-0.6 x 2-5 Filaments | Rod: 0.8 x 3-22 Filaments | Rod: 0.5 x 3-5, Filaments | Rod: 0.2-0.6 x 2- 4 Filaments | Short rod: 0.1-0.15 x 0.6-1.2, Aggregates |
| Potential temperature range C° | 25-50 | 37-39 | 15-55 | 25-45 | 15-50 | 20-45 | 15-45 | 36-45 |
| M salinity | 0.25 | 0.26 | 0.19 | 0.012 | n.a. | 0-0.3 | n.a. | 0-1.4 |
| pH range | 6.6-6.8 | 6.9-7.2 | 6.5-8.5 | 7.0-9.9 | 4.6-7.0 | 7 | 6.0-8.5 | 6.5-9.2 |

Table 1. General parameters associated with typical methanogenic species. As an example of a species adapted to the extreme cold, the species *Planococcus halocryophilus Or1*[74] survives and is metabolically active in temperatures as low as -15 °C.

The methane production rates for several methanogens have been determined from direct solid and ice samples in permafrost and glaciers. The range of $CH_4$ production varies from near 4.8 nmol/g-d to an estimated 850 nmol/g-d[75]. Typical metabolic pathways[76] that would involve carbon dioxide reduction and acetate fermentation often include reactions of the form:

$$CO_2 + 4H_2 \rightarrow CH_4 + 2H_2O \quad and \quad CH_3COO^- + H^+ \rightarrow CH_4 + CO_2 \quad . \tag{3}$$

In general, there may be several low molecular weight carboxylic acids from methanogenic extremophiles. The existence of acetate ions in the presence of water can lead to the formation of many associated acetic acid groupings involving hydrogen bonds yielding dimers[77] and hydroxyl groups[78] or cluster formations between the acetic acid subgroups[79]. These molecules have a rich vibrational spectrum that has been well studied using Raman spectroscopic techniques. Of particular interest here are the unique Raman lines for methanogenic subsurface acetic acid that can be in dry or moist Martian regolith in the presence of metallic sulfate concentrations similar to those found by the Curiosity rover. The sulfate group produces strong symmetric stretching Raman lines near 1000 $cm^{-1}$ which overlap and obscure a strong acetic acid line, however a secondary acetic acid line near 1605 $cm^{-1}$is apparent at high enough concentrations. Here we explore the threshold Raman[80] signal at 532 nm and 780 nm for acetic acid as a methanogenic byproduct mixed in JSC-1 Martian Regolith simulant with hydrated metallic sulfates blended in at ratios indicated from Curiosity rover measurements and studied by Wang[81] that would have been protected from radiation by being part of a subsurface environment.

## II. Sample Preparation and Procedure

We mixed reagent grade hydrated iron and magnesium sulfates prepared with seven different concertation levels of glacial acetic acid in ambient conditions as shown in Table 2, leading to concentrations ranging from 52 ppm-m to 653 ppm-m. Samples were mixed with JSC-1 Martian regolith simulant, with grain size distribution given in Table 3 with an optical image of the 25 mm capsule and grain distribution and whose contents are as indicated in the first column in Table 4. Each mixture was placed into a capsule and compressed to 5 MPa and dried for one week. Samples were tested at two different wavelengths to secure Raman Spectra: an inVia confocal Renishaw 532 nm, 50mW, FSR operating from 300-3500 $cm^{-1}$ and a Thermo-Scientific confocal DXR2 780 nm, 80mW, FSR from 200-3500 $cm^{-1}$ Raman Spectrometer with contact spot size of 8 microns. Sample pH measurements were done using a Jenway 3510 pH meter with an Orion 910003 calibration sample at 22.3$^o$ C.

| | | | JSC-1 86.72 g/mole | Mass Fraction | Molecules | Acetic Acid | 60.05 g/mol | ocnetratior | ppm | by mass | | |
|---|---|---|---|---|---|---|---|---|---|---|---|---|
| Sulfates | g/mole | mass g | mass g | Sulfate | N | AA 1 ppm | AA 2 ppm | AA 3 ppm | AA 4 ppm | AA 5 ppm | AA 6 ppm | pH |
| JSC-1 | 86.72 | 0.00 | 0.71 | 0.00 | 4.93E+21 | 76 | 149 | 203 | 294 | 362 | 586 | 7.72 |
| MgSO4 | 120.37 | 0.08 | 0.72 | 10.97 | 5.39E+21 | 85 | 134 | 195 | 248 | 321 | 653 | 7.84 |
| MgSO4-3H2O | 174.41 | 0.07 | 0.75 | 9.73 | 5.46E+21 | 52 | 163 | 241 | 296 | 383 | 547 | 7.51 |
| MgSO4-5H2O | 210.44 | 0.07 | 0.73 | 9.86 | 5.27E+21 | 64 | 186 | 225 | 308 | 425 | 584 | 7.45 |
| MgSO4-7H2O | 246.49 | 0.07 | 0.74 | 10.08 | 5.32E+21 | 73 | 148 | 216 | 287 | 397 | 563 | 7.73 |
| FeSO4 | 151.90 | 0.05 | 0.53 | 9.62 | 3.88E+21 | 59 | 137 | 275 | 329 | 381 | 498 | 7.81 |
| FeSO4-5H2O | 241.97 | 0.05 | 0.54 | 9.63 | 3.88E+21 | 63 | 152 | 217 | 273 | 326 | 472 | 7.03 |
| FeSO4-7H2O | 278.00 | 0.05 | 0.52 | 10.19 | 3.72E+21 | 84 | 165 | 283 | 352 | 412 | 485 | 7.14 |

Table 2. Composition of both the Fe and Mg based samples, prepared with glacial acetic acid and JSC-1 Mars simulant to give parts per million by mass.

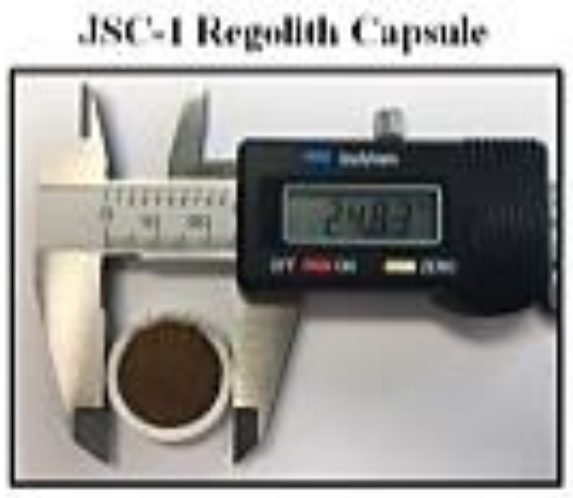

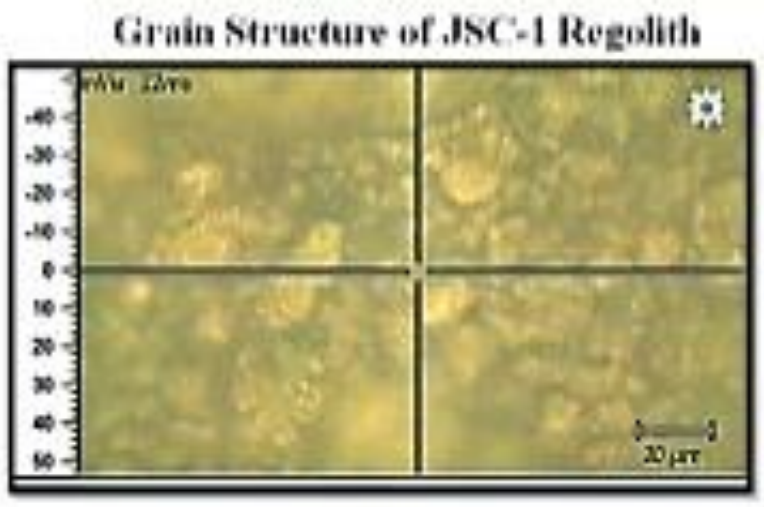

| JSC-1 | |
|---|---|
| Grain Size Distribution | |
| microns | mass % |
| 1000-450 | 21 |
| 449-250 | 30 |
| 249-150 | 24 |
| 149-53 | 19 |
| 52-5 | 5 |
| <5 | 1 |

Table 3. Image of Regolith capsule about 25 mm in diameter and 3 mm deep, 10 X Nikon Microscope optical image of granularity structure of a sample Grain size distribution of JSC-1 Martian simulant as reported by Allen, et.al[82].

The JSC-1 Mars Regolith simulant was designed by NASA Johnson Spaceflight Center to help support scientific, engineering and research labs investigate and characterize materials that can mimic as closely as possible the features of Martian regolith. Table 4 gives a comparison of JSC-1 Martian Regolith to sample sites on Mars where data was available.

# Comparison of Bulk Martian Regolith Samples

| | | 1976 | 1976 | 1997 | 2004 | 2004 | 2012 |
|---|---|---|---|---|---|---|---|
| Chemical | wt % | Chryse Planitia | Utopia Planitia | Ares Vallis | Gusev Crater | Meridiani Planum | Gale Crater |
| | JSC Mars-1 | VL-1 | VL-2 | Pathfinder | Spirit | Opportunity | Curiosity |
| $SiO_2$ | 34 | 43 | 43 | 44 | 46.1 ± 0.9 | 45.7 ± 1.3 | 42.88 ± 0.47 |
| $Al_2O_3$ | 18 | 7.3 | 7 | 7.5 | 10.19 ± 0.69 | 9.25 ± 0.50 | 9.43 ± 0.14 |
| $TiO_2$ | 3 | 0.66 | 0.56 | 1.1 | 0.88 ± 0.19 | 1.03 ± 0.12 | 1.19 ± 0.03 |
| $Fe_2O_3$ + FeO | 12 | 18.5 | 17.8 | 16.5 | 16.3 ± 1.1 | 18.8 ± 1.2 | 19.19 ± 0.12 |
| MnO | 0.2 | n.a. | n.a. | n.a | 0.32 ± 0.03 | 0.37 ± 0.02 | 0.41 ± 0.01 |
| CaO | 4.9 | 5.9 | 5.7 | 5.6 | 6.3 ± 0.29 | 6.93 ± 0.32 | 7.28 ± 0.07 |
| MgO | 2.7 | 6 | 6 | 7 | 8.67 ± 0.60 | 7.38 ± 0.29 | 8.69 ± 0.14 |
| $K_2O$ | 0.5 | <0.15 | <0.15 | 0.3 | 0.44 ± 0.07 | 0.48 ± 0.05 | 0.49 ± 0.01 |
| $Na_2O$ | 1.9 | n.a. | n.a. | 2.1 | 3.01 ± 0.30 | 2.21 ± 0.18 | 2.72 ± 0.10 |
| $P_2O_5$ | 0.7 | n.a. | n.a. | n.a. | 0.91 ± 0.31 | 0.84 ± 0.06 | 0.94 ± 0.03 |
| $SO_3$ | n.a. | 6.6 | 8.1 | 4.9 | 5.78 ± 1.25 | 5.83 ± 1.04 | 5.45 ± 0.10 |
| Cl | n.a. | 0.7 | 0.5 | 0.5 | 0.70 ± 0.16 | 0.65 ± 0.09 | 0.69 ± 0.02 |
| $Cr_2O_3$ | n.a. | n.a. | n.a. | n.a. | 0.33 ± 0.07 | 0.41 ± 0.06 | 0.49 ± 0.02 |
| Ni µg/g | n.a. | n.a. | n.a. | n.a. | 476 ± 142 | 457 ± 97 | 446 ± 29 |
| Zn µg/g | n.a. | n.a. | n.a. | n.a. | 270 ± 90 | 309 ± 87 | 337 ± 17 |
| Br µg/g | n.a. | n.a. | n.a. | n.a. | 53 ± 46 | 100 ± 111 | 26 ± 6 |

Table 4. Composition of the raw JSC-1 Martian Regolith from JPL[83] compared to Viking Landers 1 and 2 from XRF[84], Pathfinder from APXF[85], Spirit from APXS[86,87], Opportunity[88] from APXS[89] and Curiosity[90],[91] from APXS. Errors for Gusev Crater and Gale Crater, in the Rocknest[92] site, represent average values plus or minus one standard deviation, errors for the Meridiani measurements represent analytical uncertainties[93],[94]. The Curiosity ChemCam[95] identified two different soil types: a fine grained mafic similar to the Martian dust and a locally derived coarse grained felsic[96] with typical 8% one sigma average deviations. The oxychlorine compounds[97], soluble sulfates[98] and ground ices[99] were discovered by the Phoenix lander in 2008 at Vastitas Borealis and include the chlorates, perchlorates and chlorites using the Mars Environmental Compatibility Assessment (MECA) Wet Chemistry Lab[100] and Sample Analysis on Mars (SAM)[101] instrument.

## Prominent Raman Spectral Lines

| Raman Lines | | Units | |
|---|---|---|---|
| Acetic Acid | | $cm^{-1}$ | Vibrational Mode |
| | C-H | 3020 | $sp^3$ C-H, stretch $CH_3$:H-C-H asymmetric stretch |
| | C-H | 2985 | C-H asymmetric stretch |
| | | 2940 | $sp^3$ C-H, stretch $CH_2$ :$CH_3$ stretching |
| | | 1758 | C=O stretching |
| | C=O dimer | 1668 | O-H hydrogen bonded dimer |
| | | 1630 | C=O stretching |
| | $CH_2$ | 1428 | H-C-H asymmetric swing |
| | | 1366 | H-C-H bending |
| | | 1280 | O-H swing, C-C swing |
| | $CH_2$ | 1014 | O-C-C asymmetric stretch, H-C-H symmetric swing |
| | | 894 | H-C-H symmetric swing, O-H swing |
| | | 621 | O=C-O bending, C-C stretching |
| | | 603 | O=C-O bending, C-C stretching |
| | | 446 | H-C-H symmetric swing, O-H swing |
| Sulfates | | | |
| X=Ca,Fe,Mg | $XSO_4$-N $H_2O$ | 3380 | Mg stretching vibrational modes |
| | | 1006 | 982-1052 for Mg as N=0,1,…11 |
| | $SO_4$ | 1010 | for $SO_4$ symmetric stretching mode |
| | $H_2O$ | 3300 | 3200-3480 for $H_2O$, broad line stretching mode |

Table 5. Raman Lines for Acetic Acid[102,103] and potential hydrated sulfates[104,105] combined with the JSC -1 Martian simulant.

The Raman spectra with the sulfates exhibit the strong $SO_4$ peak which splits and overall broadening dependent upon hydration states and complexity of the water interaction seen in Fig. 3 (a). as noted by Sharma[106] and Wang.[13,20,107] The sulfates show the expected decrease in intensity as concentration is changed as shown in Fig. 3 (b) and (c) with the primary peak at 1006 $cm^{-1}$ with concentrations by mass percent changing as 20%, 15%, 10%,7%, 3% and 0% to approximately match amounts observed on the surface of Mars. In addition to the Raman spectra from the standard capsules with varying sulfate, acetic acid and hydration levels we considered the estimated potential Raman signals from fragmented molecular combinations that might interfere with or obscure the threshold signals of interest. We examined fragmented components that include alkanes, peroxides, dimer pairings, sulfate groupings and alcohols that can produce background signals, several of these are shown as broadband background signals in Fig. (3). For the observed acetic acid threshold, the line near 1608 $cm^{-1}$ was most often not significantly impacted by the fragmented molecular Raman lines.

# Mars Regolith-Acetic Acid Raman Spectra and Select Group Resonances

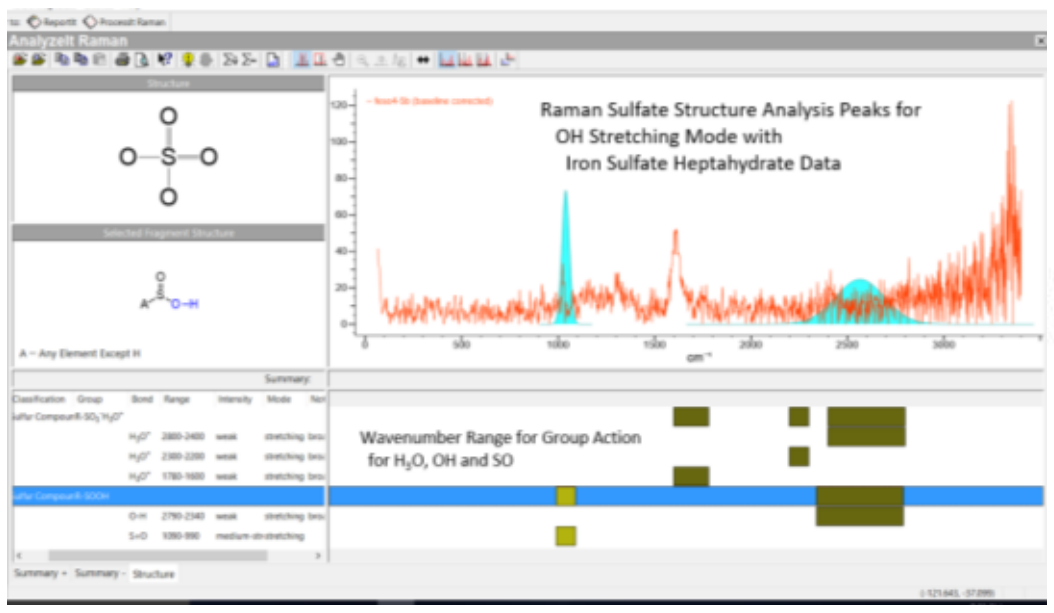

a.

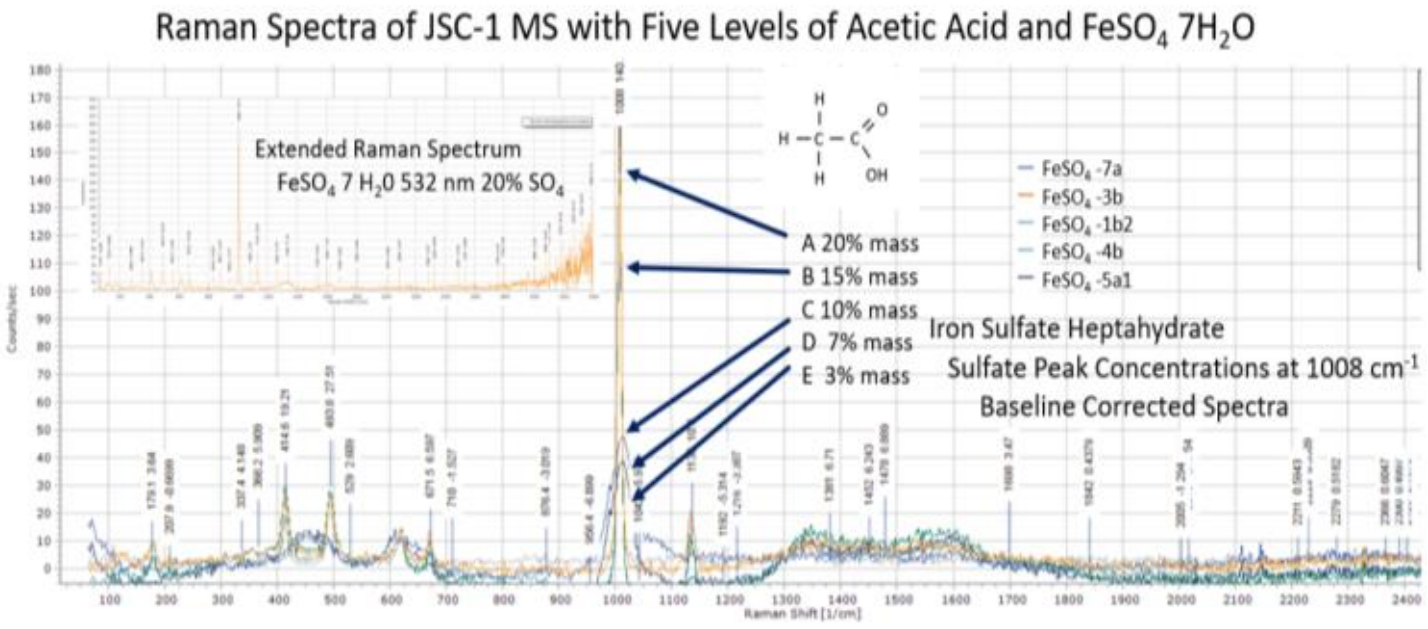

b.

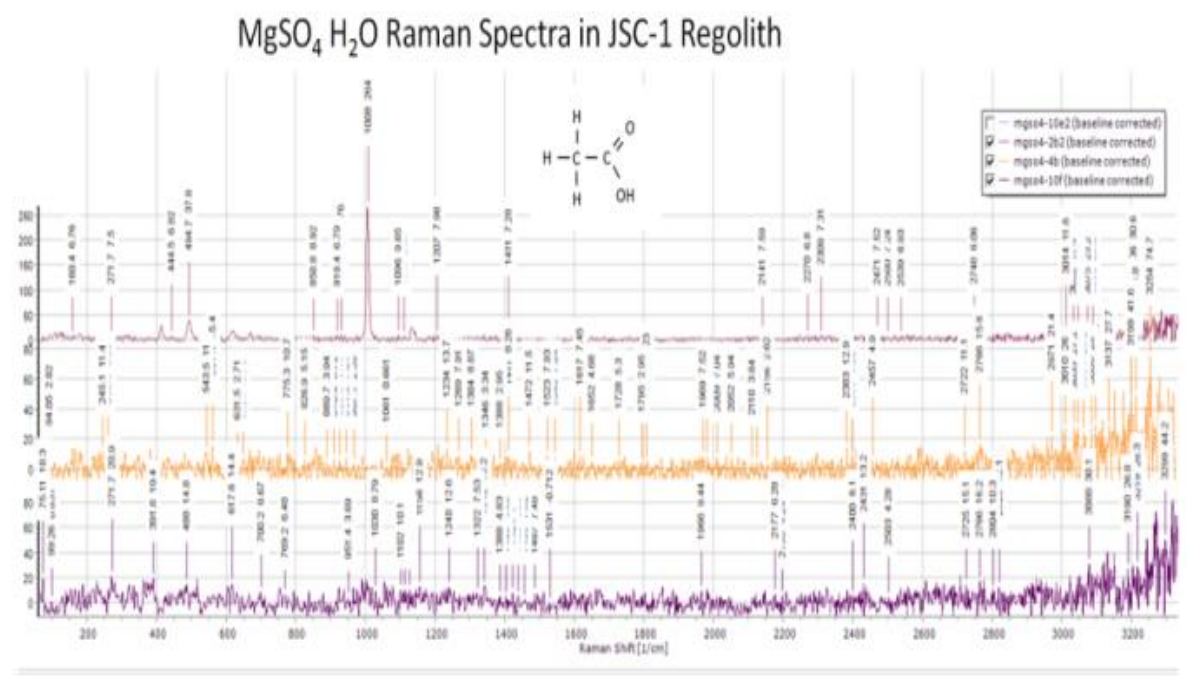

c.

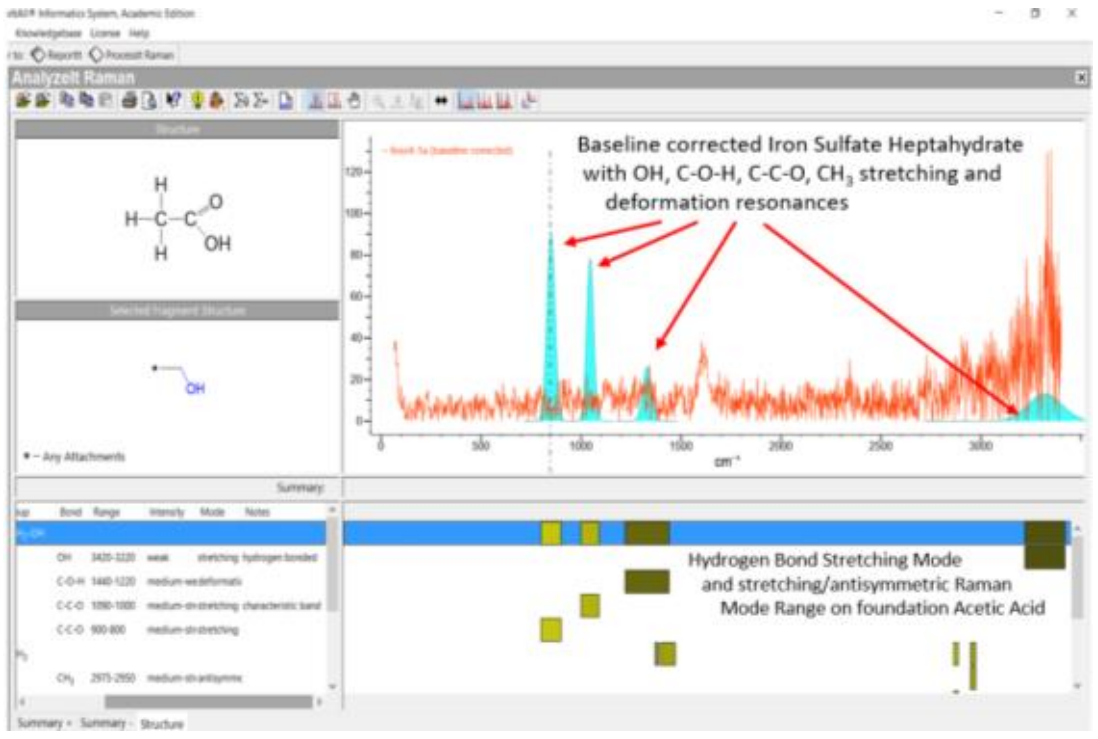

d.

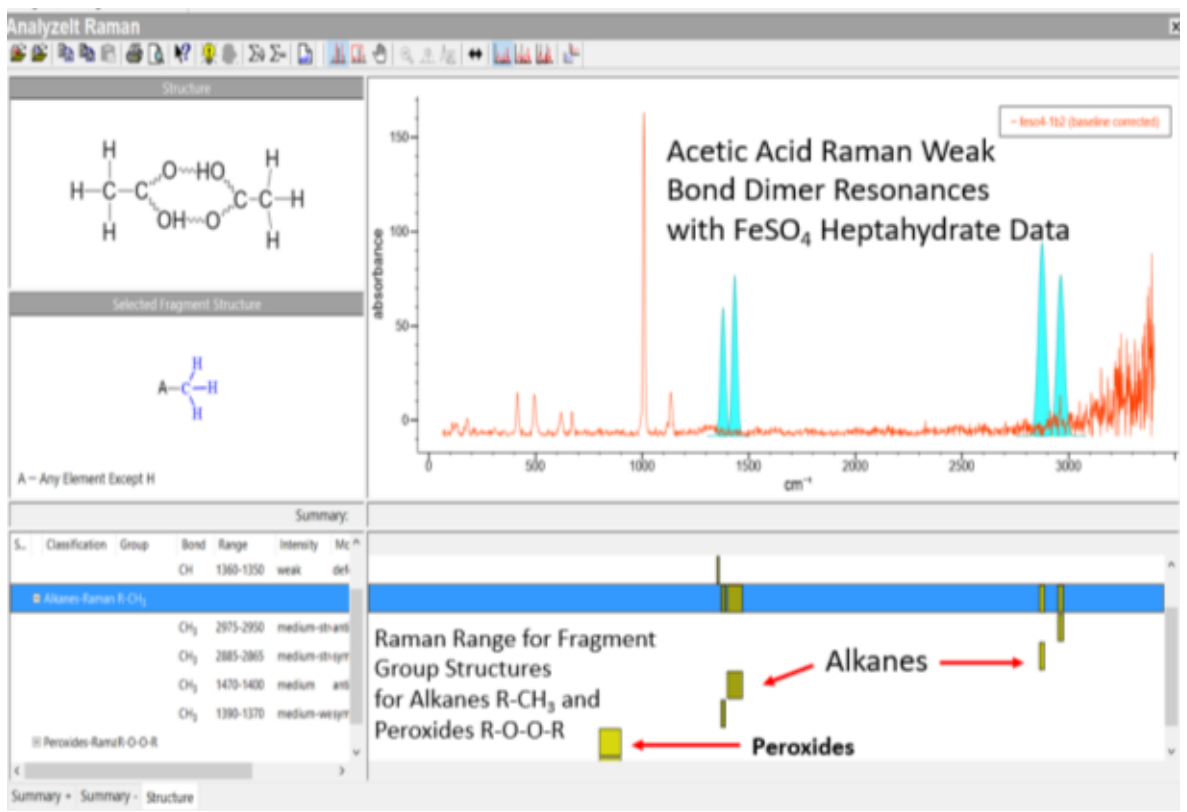

e.

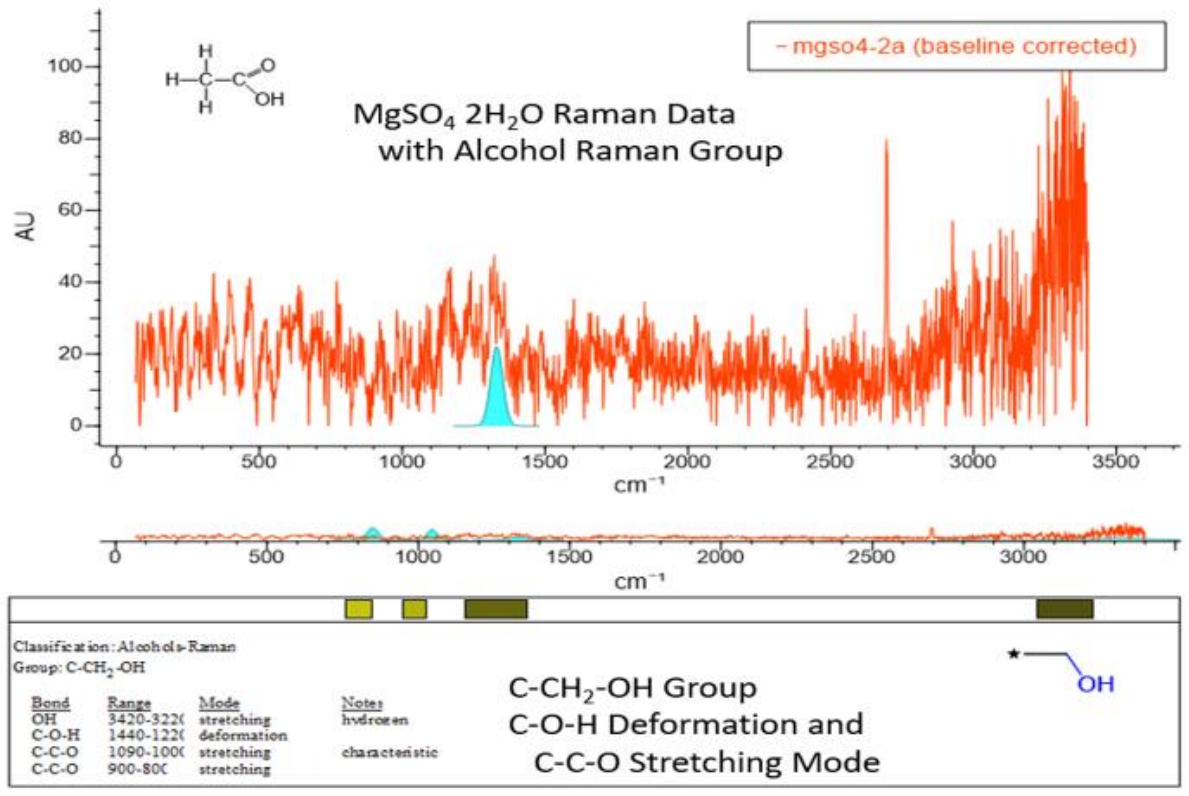

f.

Mars Regolith-Acetic Acid Raman Spectra and Select Group Resonances

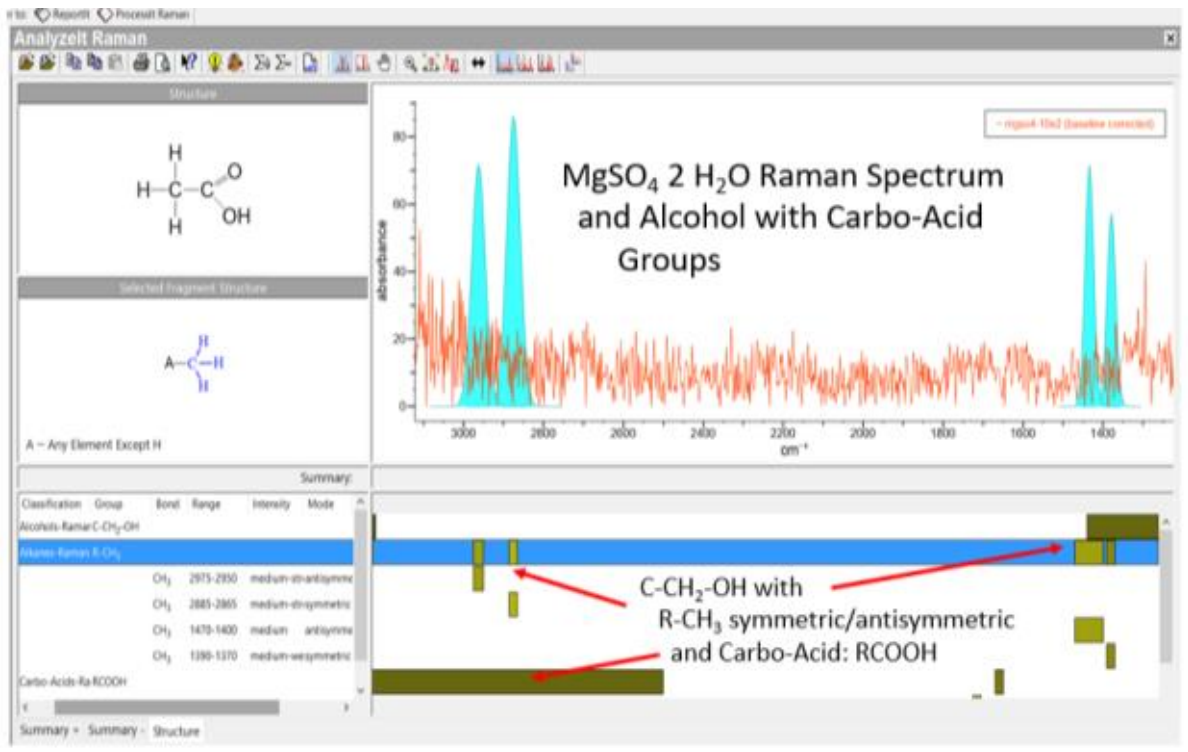

g.

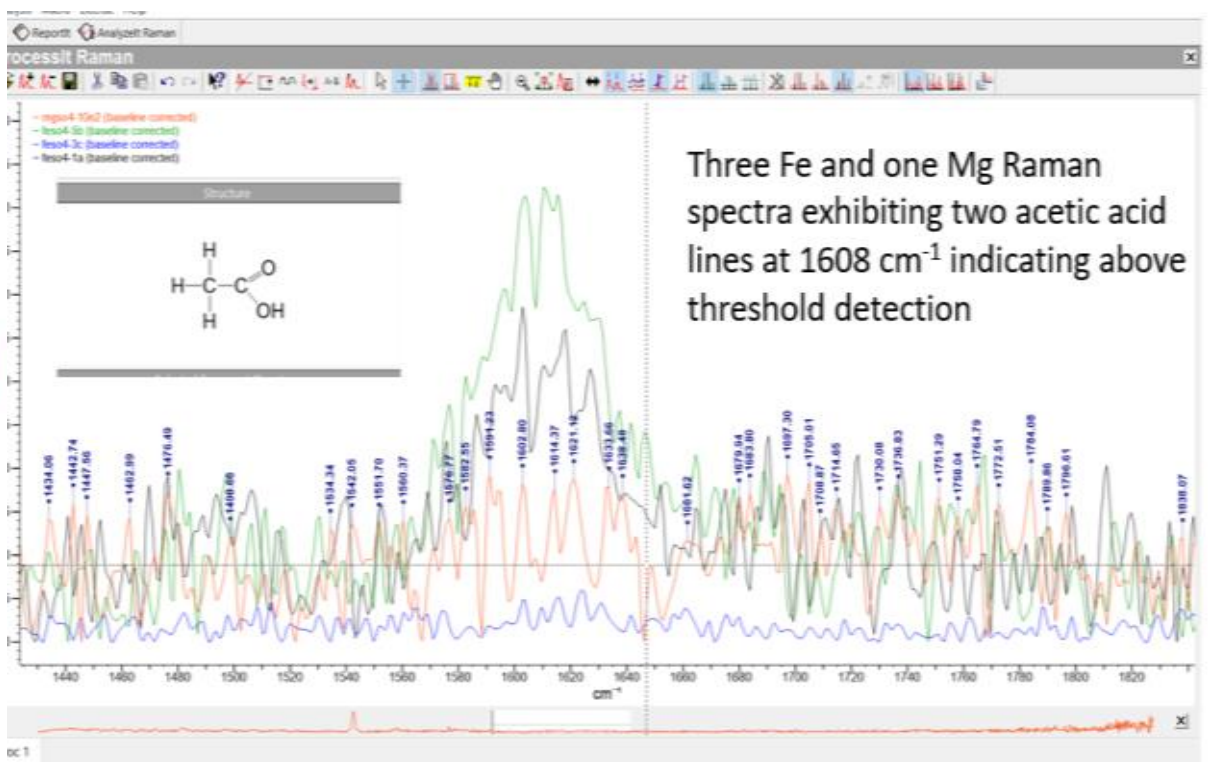

h.

Fig. 3. (a) Analytical map of Raman excitation bands due to the sulfate group combined with Fe and Mg atoms analyzed in Bio-Rad KonwItAll Raman Software[108] with fragment resonances from $H_3O$, OH and SO. (b) Hydrated $FeSO_4$ with decreasing sulfate concentrations indicated by A, B, C, D, and E, imaged using SpectraGryph[109], (c) Decrease in hydrated Mg sulfate concentrations and decrease of the 1006 cm$^{-1}$ peak, (d) analytical hydroxyl group Raman excitations in acetic acid with fragments from OH, $CH_3$, C-O-H, and C-C-O, (e) weak acetic acid dimer excitations with $CH_3$ alkanes and peroxides, (f) $MgSO_4$ $2H_2O$ with alcohol group excitations from C-$CH_2$-OH, OH, C-O-H, and C-C-O, (g) methyl group Raman excitations in acetic acid with symmetric and antisymmetric excitations present and carboxylic-acid resonances from R-COOH, (h) four levels of acetic acid concentrations showing two above threshold and two below threshold in the hydrated Fe and Mg sulfate JSC-1 Martian regolith simulant mixture.

## III.Results and Conclusions

In conclusion, we have recorded the decrease in the acetic acid hydroxyl Raman line for the 532 nm laser as the concertation was reduced when mixed with Fe and Mg sulfates from 0% to 20% by mass in the JSC-1 Martian regolith simulant. The Raman 780 nm line was not as pronounced for the same concentrations. Generally, the Raman peaks shift to longer wavelengths as the hydration is increased, approximately 6.2 cm$^{-1}$ for Mg and 5.7 cm$^{-1}$ for Fe per hydration state.

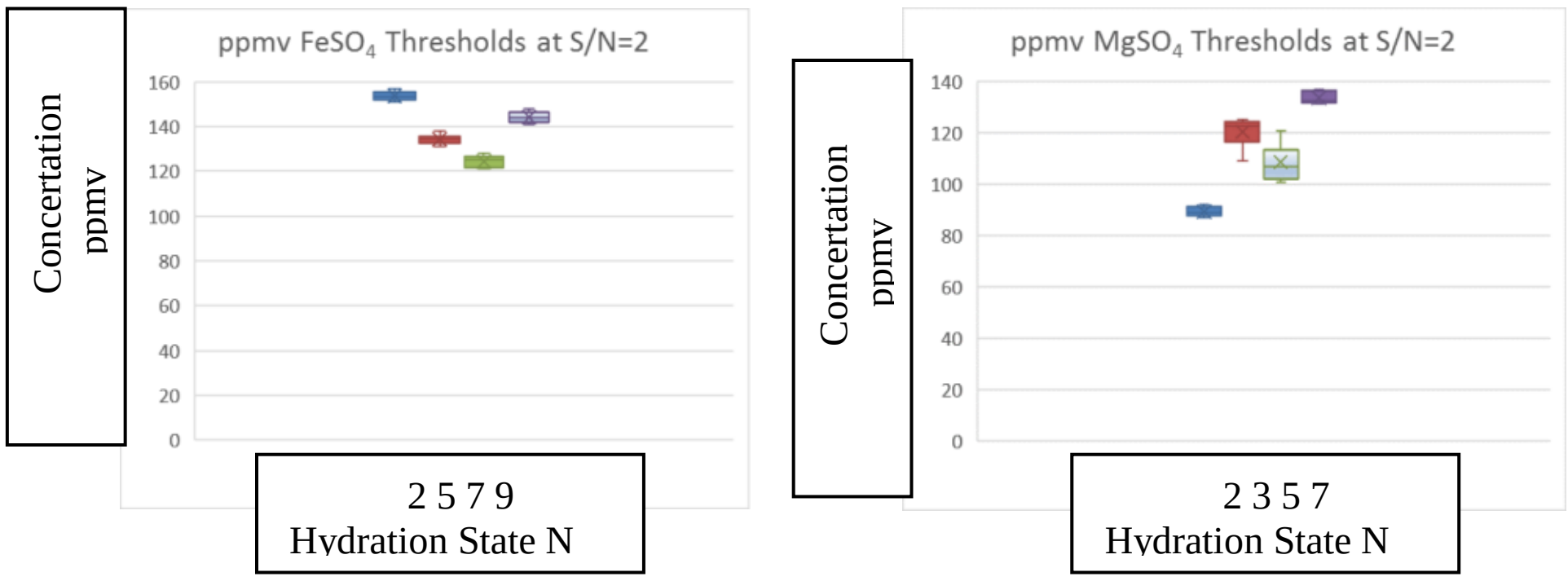

Fig. 4. Threshold values for acetic acid detection for both the iron based and magnesium based salts mixed with JSC-1 Martian regolith simulant.

We find that the minimal one sigma signal occurs near the 141-ppmv level for the Martian simulant containing the iron based sulfates and near 116-ppmv for the Mg based sulfates as shown in Fig. (4). These values are not near the ppbv level that might be required as indicated by NASA Ames[110] and Blanco[111] who demonstrate that moving to a fluorescent free Raman methodology, which has the capabilities to go below the ppm level, would be a valuable addition to the study of Martian organics[112].

## References

[1] Krasnopolsky, Vladimir A., Jean Pierre Maillard, and Tobias C. Owen. "Detection of methane in the Martian atmosphere: evidence for life?" *Icarus* 172.2 (2004): 537-547.

[2] Mumma, M. J., Villanueva, G. L., Novak, R. E., Hewagama, T., Bonev, B. P., DiSanti, M. A., & Smith, M. D. (2009). Strong release of methane on Mars in northern summer 2003. *Science*, *323*(5917), 1041-1045.

[3] Formisano, V., Atreya, S., Encrenaz, T., Ignatiev, N., & Giuranna, M. (2004). Detection of methane in the atmosphere of Mars. *Science*, *306*(5702), 1758-1761.

[4] Geminale, A., Formisano, V., & Sindoni, G. (2011). Mapping methane in Martian atmosphere with PFS-MEX data. *Planetary and Space Science*, *59*(2), 137-148.

[5] Atreya, S. K., Mahaffy, P. R., & Wong, A. S. (2007). Methane and related trace species on Mars: Origin, loss, implications for life, and habitability. *Planetary and Space Science*, *55*(3), 358-369.

[6] Lefevre, F., & Forget, F. (2009). Observed variations of methane on Mars unexplained by known atmospheric chemistry and physics. *Nature*, *460*(7256), 720-723.

[7] Zahnle, K., Freedman, R. S., & Catling, D. C. (2011). Is there methane on Mars? *Icarus*, *212*(2), 493-503.

[8] Webster, C. R., Mahaffy, P. R., Atreya, S. K., Flesch, G. J., Mischna, M. A., Meslin, P. Y., ... & Martín-Torres, J. (2015). Mars methane detection and variability at Gale crater. *Science*, *347*(6220), 415-417.

[9] Manne, J., Bui, T. Q., & Webster, C. R. (2017). Determination of foreign broadening coefficients for Methane Lines Targeted by the Tunable Laser Spectrometer (TLS) on the Mars Curiosity Rover. *Journal of Quantitative Spectroscopy and Radiative Transfer*, *191*, 59-66.

[10] Blamey, N. J., Parnell, J., McMahon, S., Mark, D. F., Tomkinson, T., Lee, M., ... & Flemming, R. L. (2015). Evidence for methane in Martian meteorites. *Nature communications*, *6*.

[11] Oze, C., & Sharma, M. (2005). Have olivine, will gas: Serpentinization and the abiogenic production of methane on Mars. *Geophysical Research Letters*, *32*(10).

[12] Morris, R. V., Klingelhoefer, G., Bernhardt, B., Schröder, C., Rodionov, D. S., De Souza, P. A., ... & Kankeleit, E. (2004). Mineralogy at Gusev Crater from the Mössbauer spectrometer on the Spirit Rover. *Science*, *305*(5685), 833-836.

[13] Fenton, L. K., Michaels, T. I., & Chojnacki, M. (2015). Late Amazonian aeolian features, gradation, wind regimes, and sediment state in the vicinity of the Mars Exploration Rover Opportunity, Meridiani Planum, Mars. *Aeolian Research*, *16*, 75-99.

[14] Morris, R. V. (2016). Martian Surface Mineralogy from Rovers with Spirit, Opportunity, and Curiosity. NASA Technical Reports: https://ntrs.nasa.gov/search.jsp?R=20160009354 .

[15] Yen, A. S., Ming, D. W., Gellert, R., Mittlefehldt, D. W., Vaniman, D. T., Thompson, L. M., ... & Arvidson, R. (2016). Similarities Across Mars: Acidic Fluids at Both Meridiani Planum and Gale Crater in the Formation of Magnesium-Nickel Sulfates.

[16] Liu, Y., & Catalano, J. G. (2016). Implications for the aqueous history of southwest Melas Chasma, Mars as revealed by interbedded hydrated sulfate and Fe/Mg-smectite deposits. *Icarus*, *271*, 283-291.

[17] Wang, A., Jolliff, B. L., Liu, Y., & Connor, K. (2016). Setting Constraints on the Nature and Origin of the two Major Hydrous Sulfates on Mars: Monohydrated and Polyhydrated Sulfates. *Journal of Geophysical Research: Planets*.

[18] Szopa, C., Millan, M., Buch, A., Belmahdi, I., Coll, P., Glavin, D. P., ... & Mahaffy, P. (2016, October). Thermal Reactivity of Organic Molecules in the Presence of Chlorates and Perchlorates and the Quest for Organics on Mars with the SAM Experiment Onboard the Curiosity Rover. In *AAS/Division for Planetary Sciences Meeting Abstracts* (Vol. 48).

[19] Millan, M., Szopa, C., Buch, A., Belmahdi, I., Coll, P., Glavin, D. P., ... & Navarro-González, R. (2016, March). Effect of the presence of chlorates and perchlorates on the pyrolysis of organic compounds: implications for measurements done with the SAM experiment onboard the Curiosity rover. In *47th Lunar and Planetary Science Conference* (p. 1418).

[20] Al Soudi, A. F., Farhat, O., Chen, F., Clark, B. C., & Schneegurt, M. A. (2016). Bacterial growth tolerance to concentrations of chlorate and perchlorate salts relevant to Mars. *International Journal of Astrobiology*, 1-7.

[21] Cockell, C. S., Catling, D. C., Davis, W. L., Snook, K., Kepner, R. L., Lee, P., & McKay, C. P. (2000). The ultraviolet environment of Mars: biological implications past, present, and future. *Icarus*, *146*(2), 343-359.

[22] Al Soudi, A. F., Farhat, O., Chen, F., Clark, B. C., & Schneegurt, M. A. (2017). Bacterial growth tolerance to concentrations of chlorate and perchlorate salts relevant to Mars. *International Journal of Astrobiology*, *16*(3), 229-235.

[23] Wadsworth, J., & Cockell, C. S. (2017). Perchlorates on Mars enhance the bacteriocidal effects of UV light. *Scientific Reports*, *7*.

[24] Board, S. S., & National Research Council. (2012). *Assessment of Planetary Protection Requirements for Spacecraft Missions to Icy Solar System Bodies*. National Academies Press.

[25] Sagan, C., & Coleman, S. (1966). Decontamination standards for Martian exploration programs. *Biology and the Exploration of Mars, Space Science Board, National Academy of Sciences, Washington, DC*, 470-481.

[26] Baqué, M., Verseux, C., Böttger, U., Rabbow, E., de Vera, J. P. P., & Billi, D. (2016). Preservation of biomarkers from cyanobacteria mixed with Marslike regolith under simulated Martian atmosphere and UV flux. *Origins of Life and Evolution of Biospheres*, *46*(2-3), 289-310.

[27] Chow, B. J., Chen, T., Zhong, Y., & Qiao, Y. (2017). Direct Formation of Structural Components Using a Martian Soil Simulant. *Scientific Reports*, *7*(1), 1151.

[28] Dundas, C.M., Bramson, A. M., Ojha, L., Wray, J. J., Mellon, M. T., Byrne, S., McEwen, A. S., Putzig, N. E., Viola, D., Sutton, S., Clark, E., Holt, J. W., (2018). Exposed subsurface ice sheets in the Martian mid-latitudes, *Science*, 359, pp 199-201.

[29] Andrew, K., Andrew, K., Thomas, M., Pesterfield, A., Lineberry, Q., & Paschal, J. (2016). Linear Spectral Shift Determination of Hydrated Metallic Sulfates in Martian Regolith Simulant JSC Mars-1. *Bulletin of the American Physical Society*, *61*.

[30] Wang, A., Freeman, J. J., Jolliff, B. L., & Chou, I. M. (2006). Sulfates on Mars: A systematic Raman spectroscopic study of hydration states of magnesium sulfates. *Geochimica et Cosmochimica Acta*, *70*(24), 6118-6135.

[31] Wang, A., & Ling, Z. C. (2011). Ferric sulfates on Mars: A combined mission data analysis of salty soils at Gusev crater and laboratory experimental investigations. *Journal of Geophysical Research: Planets*, *116*(E7).

[32] Roe, C., Broder, B., Kintzel, E., Andrew, K., Palmquest, S., & Thomas, M. (2015). Characterization of Simulated Martian Nanocomposites with Alkali Perchlorate Salts. *Bulletin of the American Physical Society, 60*.

[33] Abbey, W. J., Bhartia, R., Beegle, L. W., DeFlores, L., Paez, V., Sijapati, K., ... & Reid, R. (2017). Deep UV Raman spectroscopy for planetary exploration: The search for in situ organics. *Icarus*, *290*, 201-214.

[34] Andrew, K., Andrew, K., Thomas, M., Linebery, Q., Pesterfield, A., Womble, P. (2016). Characterization of an Extremophile Based Chlorophyll Raman Threshold Signal in Martian Regolith Simulant as a Potential Astrobiological Agent. *Bulletin of the American Physical Society, 61*.

[35] Hooijschuur, J. H., Verkaaik, M. F. C., Davies, G. R., & Ariese, F. (2016). Will Raman meet bacteria on Mars? An overview of the optimal Raman spectroscopic techniques for carotenoid biomarkers detection on mineral backgrounds. *Netherlands Journal of Geosciences*, *95*(2), 141-151.

[36] Böttger, U., de Vera, J. P., Fritz, J., Weber, I., Hübers, H. W., & Schulze-Makuch, D. (2012). Optimizing the detection of carotene in cyanobacteria in a Martian regolith analogue with a Raman spectrometer for the ExoMars mission. *Planetary and Space Science*, *60*(1), 356-362.

[37] Jehlička, J., Culka, A., & Košek, F. (2017). Obtaining Raman spectra of minerals and carbonaceous matter using a portable sequentially shifted excitation Raman spectrometer–a few examples. *Journal of Raman Spectroscopy*.

[38] Koga, T., & Naraoka, H. (2017). A new family of extraterrestrial amino acids in the Murchison meteorite. *Scientific Reports*, *7*.

[39] Pizzarello, S., & Yarnes, C. T. (2016). Enantiomeric excesses of chiral amines in ammonia-rich carbonaceous meteorites. *Earth and Planetary Science Letters*, *443*, 176-184.

[40] Oba, Y., & Naraoka, H. (2006). Carbon isotopic composition of acetic acid generated by hydrous pyrolysis of macromolecular organic matter from the Murchison meteorite. *Meteoritics & Planetary Science*, *41*(8), 1175-1181.

[41] Stevens, A. H., Patel, M. R., & Lewis, S. R. (2017). Modelled isotopic fractionation and transient diffusive release of methane from potential subsurface sources on Mars. *Icarus*, *281*, 240-247.

[42] Hu, R., Bloom, A. A., Gao, P., Miller, C. E., & Yung, Y. L. (2016). Hypotheses for near-surface exchange of methane on Mars. *Astrobiology*, *16*(7), 539-550.

[43] Lefevre, F., & Forget, F. (2009). Observed variations of methane on Mars unexplained by known atmospheric chemistry and physics. *Nature*, *460*(7256), 720-723.

[44]Moores, J. E., & Schuerger, A. C. (2012). UV degradation of accreted organics on Mars: IDP longevity, surface reservoir of organics, and relevance to the detection of methane in the atmosphere. *Journal of Geophysical Research: Planets*, *117*(E8).

[45] Michalski, J. R., & Niles, P. B. (2010). Deep crustal carbonate rocks exposed by meteor impact on Mars. *Nature Geoscience*, *3*(11), 751-755.

[46] Andrew, K., Steinfelds, E., Andrew, K., Thomas, M., Pesterfield, A., & Lineberry, Q. (2017). An Atmospheric Column Model Coupled to Surface Adsorption for Martian Methane Production in Gale Crater Using JSC-1 Martian Simulant with Metallic Sulfates. *Bulletin of the American Physical Society 62*.

[47] Moores, J. E., Smith, C. L., & Schuerger, A. C. (2017). UV production of methane from surface and sedimenting IDPs on Mars in light of REMS data and with insights for TGO. *Planetary and Space Science*.

[48]Lammer, H., Chassefière, E., Karatekin, Ö., Morschhauser, A., Niles, P. B., Mousis, O., ... & Grott, M. (2013). Outgassing history and escape of the Martian atmosphere and water inventory. *Space Science Reviews, 174*(1-4), 113-154.

[49] Greeley, R., & Schneid, B. (1991). Magma generation on Mars- Amounts, rates, and comparisons with earth, moon, and Venus. *Science*, *254*(5034), 996-998.

[50] Black, B. A., & Manga, M. (2016). The eruptibility of magmas at Tharsis and Syrtis Major on Mars. *Journal of Geophysical Research: Planets*, *121*(6), 944-964.

[51] Lillis, R. J., Dufek, J., Kiefer, W. S., Black, B. A., Manga, M., Richardson, J. A., & Bleacher, J. E. (2015). The Syrtis Major volcano, Mars: A multidisciplinary approach to interpreting its magmatic evolution and structural development. *Journal of Geophysical Research: Planets*, *120*(9), 1476-1496.

[52] Hauber, E., Brož, P., Jagert, F., Jodłowski, P., & Platz, T. (2011). Very recent and wide-spread basaltic volcanism on Mars. *Geophysical Research Letters*, *38*(10).

[53] Forget, F., Hourdin, F., Fournier, R., Hourdin, C., Talagrand, O., Collins, M., ... & Huot, J. P. (1999). Improved general circulation models of the Martian atmosphere from the surface to above 80 km. *Journal of Geophysical Research: Planets*, *104*(E10), 24155-24175.

[54] Millour, E., Forget, F., Spiga, A., Navarro, T., Madeleine, J. B., Montabone, L.,…&Lopez-Valverde, M. A. (2015, September). The Mars Climate Database (MCD version 5.2). *European Planetary Science Congress*, Vol. 10, pp. EPSC2015-438, EPSC

[55] Madeleine, J-B., F. Forget, E. Millour, L. Montabone, and M. J. Wolff. "Revisiting the radiative impact of dust on Mars using the LMD Global Climate Model." *Journal of Geophysical Research: Planets* 116, no. E11 (2011).

[56] Forget, François, Frédéric Hourdin, and Olivier Talagrand. "CO 2 snowfall on Mars: Simulation with a general circulation model." *Icarus* 131, no. 2 (1998): 302-316.

[57] Navarro, Thomas, J-B. Madeleine, François Forget, Aymeric Spiga, Ehouarn Millour, Franck Montmessin, and Anni Määttänen. "Global climate modeling of the Martian water cycle with improved microphysics and radiatively active water ice clouds." *Journal of Geophysical Research: Planets* 119, no. 7 (2014): 1479-1495.

[58] Colaïtis, Arnaud, Aymeric Spiga, Frédéric Hourdin, Catherine Rio, François Forget, and Ehouarn Millour. "A thermal plume model for the Martian convective boundary layer." *Journal of Geophysical Research: Planets* 118, no. 7 (2013): 1468-1487.

[59] González-Galindo, F., F. Forget, M. A. López-Valverde, M. Angelats i Coll, and E. Millour. "A ground-to-exosphere Martian general circulation model: 1. Seasonal, diurnal, and solar cycle variation of thermospheric temperatures." *Journal of Geophysical Research: Planets* 114, no. E4 (2009).

[60] Lefèvre, Franck, Jean-Loup Bertaux, R. Todd Clancy, Thérèse Encrenaz, Kelly Fast, François Forget, Sébastien Lebonnois, Franck Montmessin, and Séverine Perrier. "Heterogeneous chemistry in the atmosphere of Mars." *Nature* 454, no. 7207 (2008): 971-975.

[61] Seager, S., Bains, W., & Petkowski, J. J. (2016). Toward a list of molecules as potential biosignature gases for the search for life on exoplanets and applications to terrestrial biochemistry. *Astrobiology*, *16*(6), 465-485.

[62] Wynn-Williams, D. D., and H. G. M. Edwards. "Proximal analysis of regolith habitats and protective biomolecules in situ by laser Raman spectroscopy: overview of terrestrial Antarctic habitats and Mars analogs." *Icarus* 144.2 (2000): 486-503.

[63] Serrano, Paloma, et al. "Single-cell analysis of the methanogenic archaeon Methanosarcina soligelidi from Siberian permafrost by means of confocal Raman microspectrocopy for astrobiological research." *Planetary and Space Science* 98 (2014): 191-197.

[64] Miller, R. V., & Whyte, L. (Eds.). (2011). *Polar Microbiology: life in a deep freeze*. American Society for Microbiology Press.

[65] Kotelnikova, S., Macario, A. J., & Pedersen, K. (1998). Methanobacterium subterraneum sp. nov., a new alkaliphilic, eurythermic and halotolerant methanogen isolated from deep granitic groundwater. *International Journal of Systematic and Evolutionary Microbiology*, *48*(2), 357-367, includes a more detailed table.

[66] Sinha, N., & Kral, T. A. (2015). Stable carbon isotope fractionation by methanogens growing on different Mars regolith analogs. *Planetary and Space Science*, *112*, 35-41.

[67] Fabry, S., Lang, J., Niermann, T., Vingron, M., & Hensel, R. (1989). Nucleotide sequence of the glyceraldehyde-3-phosphate dehydrogenase gene from the mesophilic methanogenic

archaebacteria Methanobacterium bryantii and Methanobacterium formicicum. *The FEBS Journal*, *179*(2), 405-413.

[68] Souillard, N., Magot, M., Possot, O., & Sibold, L. (1988). Nucleotide sequence of regions homologous tonifH (nitrogenase Fe protein) from the nitrogen-fixing archaebacterial Methanococcus thermolithotrophicus and Methanobacterium ivanovii: Evolutionary implications. *Journal of molecular evolution*, *27*(1), 65-76.

[69] Worakit, S., Boone, D. R., Mah, R. A., Abdel-Samie, M. E., & El-Halwagi, M. M. (1986). Methanobacterium alcaliphilum sp. nov., an H2-utilizing methanogen that grows at high pH values. *International Journal of Systematic and Evolutionary Microbiology, 36*(3), 380-382.

[70] Patel, G. B., Sprott, G. D., & Fein, J. E. (1990). Isolation and characterization of Methanobacterium espanolae sp. nov., a mesophilic, moderately acidiphilic methanogen. *International Journal of Systematic and Evolutionary Microbiology*, *40*(1), 12-18.

[71] Zellner, G., Bleicher, K., Braun, E., Kneifel, H., Tindall, B. J., de Macario, E. C., & Winter, J. (1988). Characterization of a new mesophilic, secondary alcohol-utilizing methanogen, Methanobacterium palustre spec. nov. from a peat bog. *Archives of microbiology*, *151*(1), 1-9.

[72] König, H. (1984). Isolation and characterization of Methanobacterium uliginosum sp. nov. from a marshy soil. *Canadian journal of microbiology*, *30*(12), 1477-1481.

[73] Kotelnikova, S., Macario, A. J., & Pedersen, K. (1998). Methanobacterium subterraneum sp. nov., a new alkaliphilic, eurythermic and halotolerant methanogen isolated from deep granitic groundwater. *International Journal of Systematic and Evolutionary Microbiology*, *48*(2), 357-367.

[74] Mykytczuk, N. C., Foote, S. J., Omelon, C. R., Southam, G., Greer, C. W., & Whyte, L. G. (2013). Bacterial growth at− 15 C; molecular insights from the permafrost bacterium Planococcus halocryophilus Or1. *The ISME journal*, *7*(6), 1211.

[75] Aschenbach, K., Conrad, R., Řeháková, K., Doležal, J., Janatková, K., & Angel, R. (2013). Methanogens at the top of the world: occurrence and potential activity of methanogens in newly deglaciated soils in high-altitude cold deserts in the Western Himalayas. *Frontiers in microbiology*, *4*.

[76] Shigematsu, T., Tang, Y., Kobayashi, T., Kawaguchi, H., Morimura, S., & Kida, K. (2004). Effect of dilution rate on metabolic pathway shift between aceticlastic and nonaceticlastic methanogenesis in chemostat cultivation. *Applied and environmental microbiology*, *70*(7), 4048-4052.

[77] Kosugi, K., Nakabayashi, T., & Nishi, N. (1998). Low-frequency Raman spectra of crystalline and liquid acetic acid and its mixtures with water.: Is the liquid dominated by hydrogen-bonded cyclic dimers?. *Chemical physics letters*, *291*(3), 253-261.

[78] D'Amico, F., Bencivenga, F., Gessini, A., Principi, E., Cucini, R., & Masciovecchio, C. (2012). Investigation of acetic acid hydration shell formation through Raman spectra line-shape analysis. *The Journal of Physical Chemistry B*, *116*(44), 13219-13227.

[79] Nakabayashi, T., Kosugi, K., & Nishi, N. (1999). Liquid structure of acetic acid studied by Raman spectroscopy and ab initio molecular orbital calculations. *The Journal of Physical Chemistry A*, *103*(43), 8595-8603.

[80] Ellery, A., Wynn-Williams, D., Parnell, J., Edwards, H. G., & Dickensheets, D. (2004). The role of Raman spectroscopy as an astrobiological tool in the exploration of Mars. *Journal of Raman Spectroscopy*, *35*(6), 441-457.

[81] Wang ibid Ref. 21.

[82] Allen, C. C., Jager, K. M., Morris, R. V., Lindstrom, D. J., Lindtsrom, M. M., & Lockwood, J. P. (1998). Martian soil simulant available for scientific, educational study. *Eos, Transactions American Geophysical Union*, *79*(34), 405-409

[83] Allen ibid Ref. 68.

[84] Banin, A., Clark, B. C., & Wänke, H. (1992). Surface chemistry and mineralogy. *Mars*, 594-625.

[85] Rieder, R., Economou, T., Wänke, H., Turkevich, A., Crisp, J., Brückner, J., ... & McSween, H. Y. (1997). The chemical composition of Martian soil and rocks returned by the mobile alpha proton X-ray spectrometer: Preliminary results from the X-ray mode. *Science*, *278*(5344), 1771-1774.

[86] Squyres, S. W., Arvidson, R. E., Bell, J. F., Brückner, J., Cabrol, N. A., Calvin, W., ... & Des Marais, D. J. (2004). The Spirit rover's Athena science investigation at Gusev crater, Mars. *science*, *305*(5685), 794-799.

[87] Rieder, R., Gellert, R., Anderson, R. C., Brückner, J., Clark, B. C., Dreibus, G., ... & Squyres, S. W. (2004). Chemistry of rocks and soils at Meridiani Planum from the Alpha Particle X-ray Spectrometer. *Science*, *306*(5702), 1746-1749.

[88] Squyres, S. W., Arvidson, R. E., Bell, J. F., Brückner, J., Cabrol, N. A., Calvin, W., ... & Des Marais, D. J. (2004). The Opportunity Rover's Athena science investigation at Meridiani Planum, Mars. *science*, *306*(5702), 1698-1703.

[89] Rieder, R., Gellert, R., Anderson, R. C., Brückner, J., Clark, B. C., Dreibus, G., ... & Squyres, S. W. (2004). Chemistry of rocks and soils at Meridiani Planum from the Alpha Particle X-ray Spectrometer. *Science*, *306*(5702), 1746-1749.

[90] Leshin, L. A., Mahaffy, P. R., Webster, C. R., Cabane, M., Coll, P., Conrad, P. G., ... & Eigenbrode, J. L. (2013). Volatile, isotope, and organic analysis of martian fines with the Mars Curiosity rover. *Science*, *341*(6153), 1238937.

[91] Thompson, L. M., Schmidt, M. E., Gellert, R., & Spray, J. G. (2016). APXS Compositional trends along Curiosity's traverse, Gale Crater, Mars: Implications for crustal composition, sedimentary provenance, diagenesis and alteration. *Lunar Planet. Sci. Con. 47 Abstract, 2709*.

[92] Blake, D. F., Morris, R. V., Kocurek, G., Morrison, S. M., Downs, R. T., Bish, D., ... & Madsen, M. B. (2013). Curiosity at Gale crater, Mars: Characterization and analysis of the Rocknest sand shadow. *Science*, *341*(6153), 1239505.

[93] Vaniman, D. T., Bish, D. L., Ming, D. W., Bristow, T. F., Morris, R. V., Blake, D. F., ... & Rice, M. (2013). Mineralogy of a mudstone at Yellowknife Bay, Gale crater, Mars. *Science*, 1243480.

[94] Grotzinger, J. P., Blake, D. F., Crisp, J., Edgett, K. S., Gellert, R., Gomez-Elvira, J., ... & Meyer, M. (2013, March). Mars Science Laboratory: First 100 sols of geologic and geochemical exploration from Bradbury Landing to Glenelg. In *Lunar and Planetary Science Conference* (Vol. 44, p. 1259).

[95] Maurice, S., Clegg, S. M., Wiens, R. C., Gasnault, O., Rapin, W., Forni, O., ... & Nachon, M. (2016). ChemCam activities and discoveries during the nominal mission of the Mars Science Laboratory in Gale crater, Mars. *Journal of Analytical Atomic Spectrometry*, *31*(4), 863-889.

[96] Meslin, P. Y., Gasnault, O., Forni, O., Schröder, S., Cousin, A., Berger, G., ... & Le Mouélic, S. (2013). Soil diversity and hydration as observed by ChemCam at Gale Crater, Mars. *Science*, *341*(6153), 1238670.

[97] Hecht, M. H., Kounaves, S. P., Quinn, R. C., West, S. J., Young, S. M. M., Ming, D. W., ... & DeFlores, L. P. (2009). Detection of perchlorate and the soluble chemistry of martian soil at the Phoenix lander site. *Science*, *325*(5936), 64-67.

[98] Kounaves, S. P., Hecht, M. H., Kapit, J., Quinn, R. C., Catling, D. C., Clark, B. C., ... & Shusterman, J. (2010). Soluble sulfate in the martian soil at the Phoenix landing site. *Geophysical Research Letters*, *37*(9).

[99] Mellon, M. T., Arvidson, R. E., Sizemore, H. G., Searls, M. L., Blaney, D. L., Cull, S., ... & Markiewicz, W. J. (2009). Ground ice at the Phoenix landing site: Stability state and origin. *Journal of Geophysical Research: Planets*, *114*(E1).

[100] Kounaves, S. P., Hecht, M. H., Kapit, J., Gospodinova, K., DeFlores, L., Quinn, R. C., ... & Ming, D. W. (2010). Wet Chemistry experiments on the 2007 Phoenix Mars Scout Lander mission: Data analysis and results. *Journal of Geophysical Research: Planets*, *115*(E1).

[101] Kounaves, S. P., Hecht, M. H., West, S. J., Morookian, J. M., Young, S. M., Quinn, R., ... & Fisher, A. (2009). The MECA wet chemistry laboratory on the 2007 Phoenix Mars Scout lander. *Journal of Geophysical Research: Planets*, *114*(E3).

[102] Wan, F., Du, L., Chen, W., Wang, P., Wang, J., & Shi, H. (2017). A Novel Method to Directly Analyze Dissolved Acetic Acid in Transformer Oil without Extraction Using Raman Spectroscopy. *Energies*, *10*(7), 967.

[103] Karthikeyan, N., Prince, J. J., Ramalingam, S., & Periandy, S. (2014). Vibrational spectroscopic [FT-IR, FT-Raman] investigation on (2, 4, 5-Trichlorophenoxy) Acetic acid using computational [HF and DFT] analysis. *Spectrochimica Acta Part A: Molecular and Biomolecular Spectroscopy*, *124*, 165-177.

[104] Lopez-Reyes, G., Sobron, P., Lefebvre, C., & Rull, F. (2014). Multivariate analysis of Raman spectra for the identification of sulfates: Implications for ExoMars. *American Mineralogist*, *99*(8-9), 1570-1579.

[105] Rapin, W., Meslin, P. Y., Maurice, S., Vaniman, D., Nachon, M., Mangold, N., ... & Martínez, G. M. (2016). Hydration state of calcium sulfates in Gale crater, Mars: Identification of bassanite veins. *Earth and Planetary Science Letters*, *452*, 197-205.

[106] Sharma, S. K., Chio, C. H., & Muenow, D. W. (2006, March). Raman spectroscopic investigation of ferrous sulfate hydrates. In *37th Annual Lunar and Planetary Science Conference* (Vol. 37).

[107] Wang, A., & Zhou, Y. (2014, March). Rates of Al-, Fe-, Mg-, Ca-Sulfates Dehydration Under Mars Relevant Conditions. In *Lunar and Planetary Science Conference* (Vol. 45, p. 2614).

[108] Bio-Rad, KnowItAll Raman Analysis, Chemical Group Functionality and Resonances, http://www.bio-rad.com/en-us/spectroscopy .

[109] Menges, F., SpectraGryph Software, *Friedrich.menges@effemm2.de Web: http://spectroscopy.ninja, 2016*.

[110] Wilhelm, M.B., Sansano, A., Sanz-Arranz, J.A., , Sobron, P., , Rull, F., Davila, A. F., Critical Assessment of Biosignature Detection with Raman Spectroscopy on Biologically Lean Soils, Astrobiology Science Conference 2017 (LPI Contrib. No. 1965)

[111] Blanco, Y., Gallardo-Carreno, I., Ruiz-Bermejo, M., Puente-Sánchez, F., Cavalcante-Silva, E., Quesada, A., ... & Parro, V. (2017). Critical Assessment of Analytical Techniques in the Search for Biomarkers on Mars: A Mummified Microbial Mat from Antarctica as a Best-Case Scenario. *Astrobiology*, *17*(10), 984-996.

[112] Tang, S., Chen, B., McKay, C. P., Navarro-Gonzálezv, R., & Wang, A. X. (2016). Detection of trace organics in Martian soil analogs using fluorescence-free surface enhanced 1064-nm Raman Spectroscopy. *Optics express*, *24*(19), 22104-22109.